\documentclass[aps,prl,twocolumn,letterpaper,superscriptaddress,showpacs,10pt,nofootinbib]{revtex4-1}
\usepackage{graphicx,xcolor}
\usepackage{amsmath}
\usepackage{subfigure}
\usepackage{CJK}
\usepackage[T1]{fontenc}
\usepackage{array}
\usepackage{multirow}
\usepackage{enumerate}
\usepackage{mdwlist}
\PassOptionsToPackage{hyphens}{url}\usepackage{hyperref}

\def\tx{\text}
\def\cp{c^{\dagger}}
\def\sg{\sigma}
\def\etal{\textit{et al.}}
\def\tc{$T_c$}
\def\vep{\varepsilon}
\def\bi{{\bf i}}
\def\bj{{\bf j}}

\def\tx{\text}
\def\vep{\varepsilon}
\def\ns{\notag\\}
\def\l({\left(}
\def\r){\right)}
\def\ce{\ \tx{,}}
\def\dd{d^{\dagger}}
\def\upa{\uparrow}
\def\dna{\downarrow}

\usepackage{color}

\begin{document}
\begin{CJK*}{UTF8}{bsmi}
\title{What is the valence of Mn in Ga$_{1-x}$Mn$_x$N?}
\author{Ryky Nelson}
\affiliation{Department of Physics \& Astronomy, Louisiana State University, Baton Rouge, LA 70803}
\affiliation{Center for Computation \& Technology, Louisiana State University, Baton Rouge, LA 70803}
\author{Tom Berlijn}
\affiliation{Center for Nanophase Materials Sciences and Computer 
Science and Mathematics Division, Oak Ridge National Laboratory, Oak Ridge, TN 37831}
\author{Juana Moreno}
\affiliation{Department of Physics \& Astronomy, Louisiana State University, Baton Rouge, LA 70803}
\affiliation{Center for Computation \& Technology, Louisiana State University, Baton Rouge, LA 70803}
\author{Mark Jarrell}
\affiliation{Department of Physics \& Astronomy, Louisiana State University, Baton Rouge, LA 70803}
\affiliation{Center for Computation \& Technology, Louisiana State University, Baton Rouge, LA 70803}
\author{Wei Ku(\CJKfamily{bsmi}顧威)}
\affiliation{Condensed Matter Physics and Materials Science Department,
Brookhaven National Laboratory, Upton, New York 11973}
\affiliation{Physics Department, State University of New York, Stony Brook,
New York 11790}

\date{\today}

\begin{abstract}
We investigate the current debate on the Mn valence in Ga$_{1-x}$Mn$_x$N, a diluted magnetic semiconductor (DMSs) with a potentially high Curie temperature.
From a first-principles Wannier-function analysis, we unambiguously find the Mn valence to be close to $2+$ ($d^5$), but in a mixed spin configuration with average magnetic moments of 4$\mu_B$. 
By integrating out high-energy degrees of freedom differently, we further derive for the first time from first-principles two low-energy pictures that reflect the intrinsic dual nature of the doped holes in the DMS:
1) an effective $d^4$ picture ideal for local physics, and 2) an effective $d^5$ picture suitable for extended properties.
In the latter, our results further reveal a few novel physical effects, and pave the way for future realistic studies of magnetism.
Our study not only resolves one of the outstanding key controversies of the field, but also exemplifies the general need for multiple effective descriptions to account for the rich low-energy physics in many-body systems in general.
\end{abstract}


\maketitle
\end{CJK*}
Diluted magnetic semiconductors (DMS) have attracted great interest because of their potential applications in spintronic technology 
\cite{i_zutic_04} such as nonvolatile memory \cite{g_prinz_98a,s_dassarma_01}, spin-generating solar cells \cite{i_zutic_01,b_endres_13}, 
electrical spin injection \cite{y_ohno_99}, spin-LED (light-emitting diode) \cite{s_wolf_01}, and electrically or optically controlled 
ferromagnets \cite{s_pearton_04}.
Among the DMS materials, Ga$_{1-x}$Mn$_x$N is of particular interest and increasingly studied. One of the motivations is that 
blue LED \cite{amano1986,akasaki1993,nakamura1994} technology is based on the host compound GaN. 
Ga$_{1-x}$Mn$_x$N also might be instrumental toward the realization of efficient spintronic devices  as  
Dietl \etal\ \cite{t_dietl_01a} predicted its Curie temperature (\tc) to be
above room temperature; a feature which is obviously required in order to be technologically advantageous.
However, this prediction remains far from being fulfilled as various experiments lead to controversial conclusions concerning the ferromagnetism in Ga$_{1-x}$Mn$_x$N.
Chen \etal\ \cite{di_chen_2008} detected superparamagnetism in their nanocluster Ga$_{1-x}$Mn$_x$N sample, while Zaj\k{a}c \etal\ \cite{m_zajac_01} 
and Granville \etal\ \cite{s_granville_10} report antiferromagnetic coupling between Mn ions in their sample.
Interestingly, Dhar \etal\ \cite{s_dhar_03} in their investigation observe a Heisenberg spin-glass with a transition temperature around $4.5$ K.
Observations of the desired ferromagnetic ordering on the other hand have also been reported, albeit with fiercely varying \tc's;  some 
\cite{m_overberg_01,s_stefanowicz_13} find low \tc's between $10$ K and $25$ K, while others \cite{g_thaler_02,m_reed_01} report ferromagnetism around 
room temperature or higher \cite{t_sasaki_02}.

One factor considered to be instrumental for the magnetic order and the coupling mechanism in DMS is the valence state of Mn \cite{t_jungwirth_06,c_liu_05,Sato10}.
There is no doubt that in addition to a local moment, a (Ga,Mn) substitution injects a hole into the system, but the question is: where is 
this hole located? If the hole resides mostly in the N-valence bands and is likely delocalized, resulting in a Mn valence of $2+$($d^5$).
In this case, similar to Ga$_{1-x}$Mn$_x$As systems \cite{t_dietl_01a,m_abolfath_01}, the microscopic mechanism is described by pictures 
of Zener's kinetic-exchange type \cite {c_zener_51b}, in which the coupling between local moments is mediated by valence-band itinerant carriers.
This mechanism has been examined experimentally for Ga$_{1-x}$Mn$_x$As \cite{y_ohno_99,t_hayashi_01,k_edmonds_02,k_yu_02a,rokhinson2007}.
If, on the other hand, the hole resides mostly in Mn ions, the Mn valence is $3+$($d^4$), and the magnetic coupling would be better 
described by a double-exchange mechanism \cite{c_zener_51, n_furukawa_99} mediated by impurity levels \cite{k_sato_04,Sato10}.

Despite its widely accepted importance, the Mn valence state  in Ga$_{1-x}$Mn$_x$N is still controversial.
Early experimental \cite{r_korotkov_01,r_korotkov_02,burch_08} and density functional theory (DFT) studies 
\cite{e_kulatov_02,b_sanyal_03,l_sandratskii_04,m_wierzbowska_04,j_kang_05,p_mahadevan_04} demonstrated a partially filled impurity band 
formed deeply in the band gap with a significant Mn $d$ character, suggesting a Mn$^{3+}$ ($d^4$) configuration different from the Mn$^{2+}$ ($d^5$) 
one in Ga$_{1-x}$Mn$_x$As \cite{j_schneider_87}. Later, both x-ray absorption spectroscopy (XAS) studies \cite{x_biquard_03,Titov05,w_stefanowicz_10} and optical absorption 
analysis \cite{t_graf_02,t_graf_03} also concluded a Mn valence state of $3+$ ($d^4$).
However,  other XAS studies \cite{y_soo_01,k_edmonds_04b,Hwang05} demonstrate that Mn is predominantly Mn$^{2+}$ ($d^5$).
A similar conclusion was also reached by electron spin resonance \cite{m_zajac_01} and magnetic measurements \cite{s_granville_10}.
Clearly, a resolution of the uncertainty about the Mn valence state is imperative for further progress in the understanding and engineering of 
the Ga$_{1-x}$Mn$_x$N DMS. 

In this Letter, we investigate the controversial Mn valence state in Ga$_{1-x}$Mn$_x$N.
Our first-principles Wannier-functions based analysis \cite{w_ku_02} covering the high-energy Hilbert space demonstrates unambiguously that the Mn valence is close to $2+$ ($d^5$) but with a mixed spin configuration that gives average magnetic moments of 4$\mu_B$ (\textit{not} 5$\mu_B$).
Interestingly, at the more relevant lower-energy scale, due to the proximity of N $s$ and $p$ energy levels to the Mn $d$ level, the dual nature of the doped hole can be realized.
Defining Wannier orbitals (WOs) in a narrower energy range, we show the feasibility of both the effective $d^4$ and $d^5$ descriptions, which are convenient to describe different physical aspects of Ga$_{1-x}$Mn$_x$N.
The resulting effective $d^4$ picture offers the simplest description of the local magnetic moment and the Jahn-Teller distortion while the effective $d^5$ picture is most suitable for long-range magnetic order.
Moreover, our first-principles result reveals several strong physical effects absent in previous studies.
Our study not only resolves one of the outstanding key puzzles in the field of DMSs, but also highlights the generic need for multiple effective descriptions in describing the rich low-energy physics in interacting systems in general.

We start by performing first-principles DFT calculations in a zinc blende supercell of 64 atoms (Ga$_{31}$MnN$_{32}$) within the full-potential 
linearized augmented-plane-wave method \cite{Blaha2001}.
The LDA+U approximation \cite{anisimov1993} is applied to Mn atoms with $U = 4$ eV and $J = 0.8$ eV.
We then construct WOs \cite{w_ku_02} in three different ways to effectively integrate out various degrees of freedoms, to analyze the electronic 
structure at different energy scales, and to illustrate the relevant physical effects.
As will become clear below, the use of WOs is crucial in the analysis, for example in counting the charges.

\begin{figure}[ht]
\includegraphics[width=0.5\textwidth,clip=true]{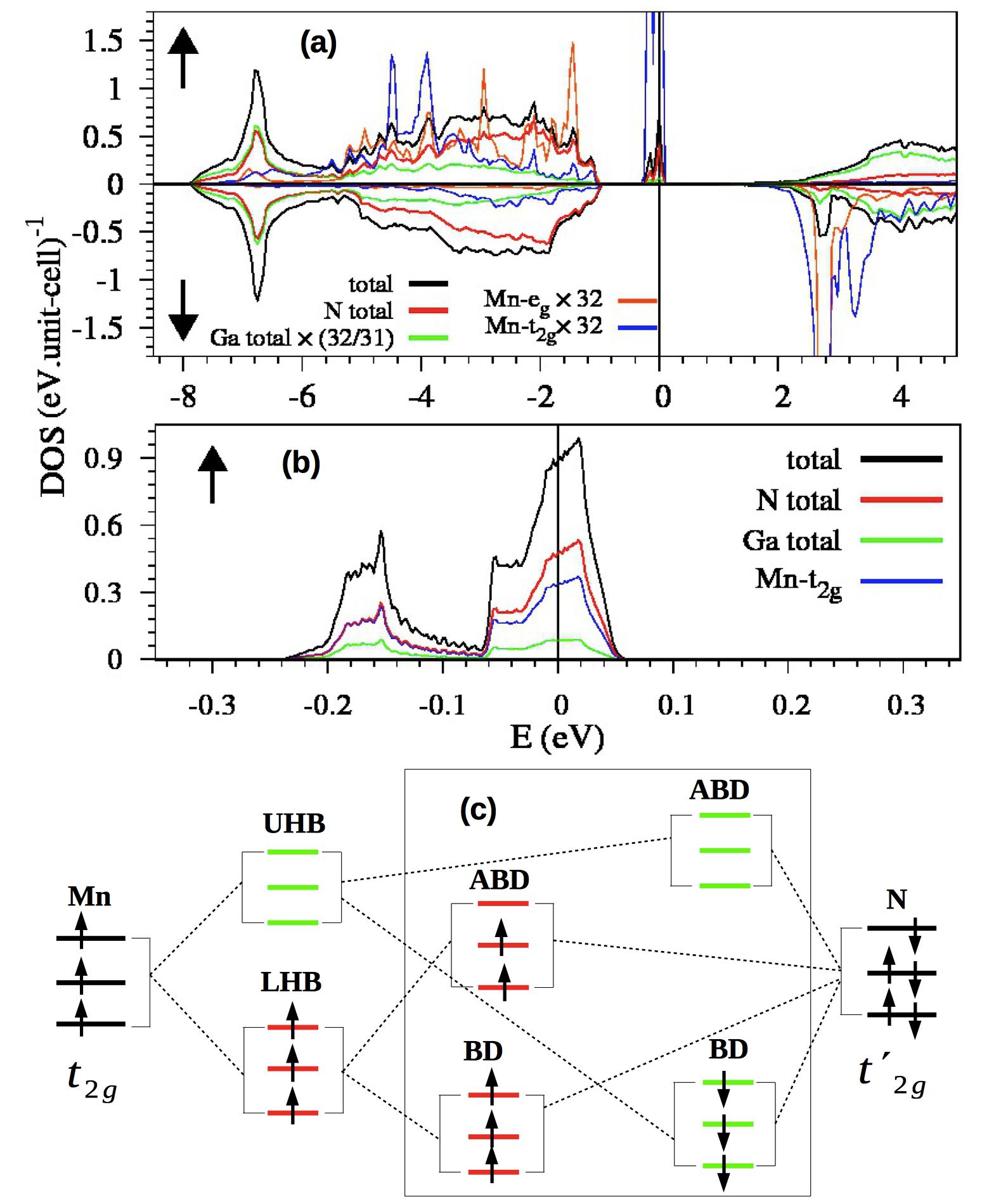}
\caption{
(color online) (a) Total and partial densities of states (DOSs) of Ga$_{31}$MnN$_{32}$ with the Fermi energy ($E_\tx{F}$) at $0$ eV.
The partial DOSs have been scaled up in units of "per atom of this kind."
(b) The DOS of the impurity bands around E$_\tx{F}$ in units of "per GaN primitive unitcell."
(c) Illustration of the hybridization of Mn-N orbitals.
Up and down arrows represent spin majority and minority, respectively.
UHB (LHB) denotes upper (lower) Hubbard bands, whereas BD (ABD) denotes bonding (antibonding).
Note that two electrons residing in the spin-majority $e_g$ orbitals are not shown here.
}
\label{fig:dos}
\end{figure}

First, to address the question on the valence state of Mn we look into the high-energy properties by analyzing the resulting density of states with N-$sp^3$, Ga-$sp^3$, and Mn-$d$ symmetries 
covering the energy range of $[-18.0, 9.0]$ eV. 
Figure \ref{fig:dos}(a) shows partially filled impurity bands lying deep in the band gap similarly to previous DFT analyses 
\cite{e_kulatov_02,b_sanyal_03,l_sandratskii_04, m_wierzbowska_04,j_kang_05, Sato10}.
Particularly, Fig. \ref{fig:dos} (b) shows that the Mn-$t_{2g}$ impurity levels are strongly hybridized with the surrounding N-$sp^3$ orbitals, 
such that the total weight in the N orbital slightly exceeds that of the Mn.
Integrating the DOS up to the Fermi energy, we find the Mn occupation to be 5.0, corresponding to the Mn valence of ${2+}$ ($d^5$).
This result is quite different from the value of 4.59 presented in a previous DFT study \cite{e_kulatov_02}, but the distinction is easily 
understandable from the fact that counting charges within an artificially chosen muffin tin around the Mn ion would necessarily miss the 
interstitial contributions.
Our WOs, on the other hand, span the entire Hilbert space up to $9$ eV and leave no unaccountable charges.

However, this seemingly clean Mn$^{2+}$ charge distribution contributes to a total spin of only 4.0$~\mu_B$ (\textit{not} 5$~\mu_B$).
Therefore, it should not (and cannot) be understood simply from the pure ionic $d^5$ configuration.
Indeed, Fig. \ref{fig:dos}(a) shows clearly that part of the 5.0 $d$ electrons resides in the spin minority channel spreading over a large energy 
range, as a consequence of the strong hybridization with the N orbitals.
A simpler visualization of this beyond-ionic configuration is given by Fig. \ref{fig:dos}(c) that summarizes the basic building blocks of the 
electronic structure. It is now clear that the Mn ion hosts part of a hole in the antibonding orbitals of the lower Hubbard bands (LHBs), and part 
of three electrons in the bonding orbitals of the upper Hubbard bands (UHBs).
Specifically, we found 0.5 electrons in the Mn spin-minority channel, and 4.5 in the majority one, giving a net moment of 4.0$~\mu_B$.
(The N orbitals that hybridize with Mn orbitals, named N-$t'_{2g}$ in Fig. \ref{fig:dos}(c), will be defined in detail below.)

Obviously now, the strong hybridization between Mn and N orbitals renders the high-energy ionic picture based on atomic orbitals completely 
inapplicable in the lower-energy sector, in which the renormalized orbitals absorb the hybridization upon integrating-out the higher-energy 
degrees of freedom. In other words, at low energy, electrons are no longer able to reside in Mn or N atomic orbitals, but only in Mn-N hybrids.
Therefore, debating the ionic valency with atomic orbitals is of no physical significance for the low-energy behavior of the system.
Instead, the physics should be described by effective or ``renormalized'' Mn and N orbitals. 

Interestingly, the proximity of the N and Mn orbital energies, which enhances the hybridization and other quantum effects, also enables the
generic possibility of multiple representations of the many-body system.
It is feasible to derive multiple low-energy effective pictures, depending on which is more convenient for describing the physical 
properties of interest. Below, we demonstrate this fundamental feature by constructing various low-energy effective WOs that correspond to 
integrating-out higher-energy degrees of freedom differently.
Specifically, we show that both effective $d^4$ and $d^5$ pictures can be derived, and both are useful for describing certain properties.

\begin{figure}[ht]
\includegraphics[width=1\linewidth,clip=true]{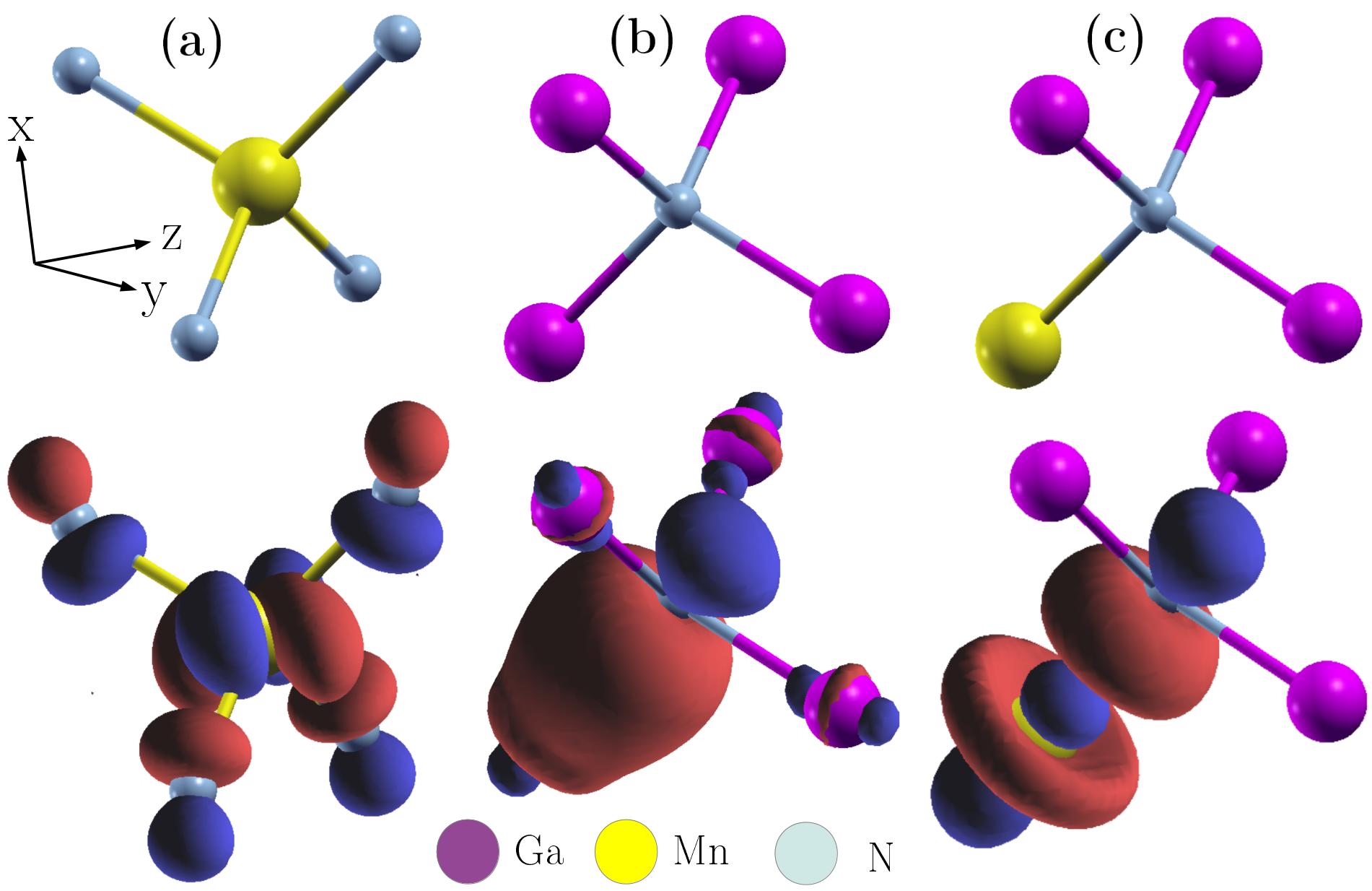}
\caption{
(color online) Illustration of the WOs used in (a) low-energy effective $d^4$ and (b),(c) the effective $d^5$ picture.
The upper panels show the local crystal structure, while the lower panels plot the isosurface of (a) $\widetilde{\tx{Mn}}-t_{2g}$, and (b)(c) $\widetilde{\tx{N}}-sp^3$ WOs at $0.07$ bohr$^{-3/2}$.
}
\label{fig:wan_orb}
\end{figure}

We start with the local properties of Ga$_{1-x}$Mn$_x$N. 
Figure \ref{fig:dos}(b) shows a $\frac{2}{3}$--filled impurity level, corresponding to two electrons residing in three degenerate ``effective'' 
$t_{2g}$ WOs.
One thus expects a strong local Jahn-Teller instability toward splitting the degeneracy into $2+1$.
Indeed, the Jahn-Teller instability has been found in previous studies \cite{x_luo_05,a_boukortt_12}.
It is easier to describe this local physics using an ``effective'' $d^4$ picture. 
Figure \ref{fig:wan_orb} (a) shows one of the effective $\widetilde{\tx{Mn}}$-$t_{2g}$ WOs corresponding to the impurity levels between [$-0.4,0.4$].
It has the symmetry of the Mn-$t_{2g}$ orbital, but with large tails in the surrounding N ions, incorporating the antibonding hybridization 
illustrated in Fig. \ref{fig:dos}(c).
It is in this effective $\widetilde{\tx{Mn}}$-$t_{2g}$ WOs that an effective $d^4$ picture is realized: A threefold degenerate WO 
hosting two electrons, which then split into $2+1$ orbitals upon orbital polarization.
(The other two electrons reside in the spin-majority effective $e_g$ WOs.)
This effective $d^4$ picture also gives a local moment of 4$\mu_B$ that is really the one fluctuating at low energy, with a form factor 
\cite{walters2009,larson2007} extending to neighboring N ions in real space.

An interesting point that emerges here is that the hybridization with Mn-$t_{2g}$ naturally splits the surrounding four N-$sp^3$ orbitals, one from 
each N ion pointing toward Mn, into a set of $3+1$ configurations.
The threefold degenerate ones have the correct signs to match each of the Mn-$t_{2g}$ orbitals: ($+,+,-,-$), ($+,-,+,-$), ($+,-,-,+$), while the fourth 
one with sign ($+,+,+,+$) does not couple to the Mn-$t_{2g}$ orbitals.
One thus can conveniently name them N-$t^\prime_{2g}$ and N-$s^\prime$ WOs centered at the Mn site.
The four tails of the WOs in Fig. \ref{fig:wan_orb} (a) give an example of one of these N-$t^\prime_{2g}$ orbitals 
which in Fig. \ref{fig:dos}(c) hybridize with Mn-$t_{2g}$. These N-$t^\prime_{2g}$ are the ones being integrated out to derive the effective $d^4$ 
picture. Note that this change of perspective is the same as that employed in the construction of the well-known Zhang-Rice singlet in the cuprate 
high-temperature superconductors \cite{f_zhang_88,MB_DownFolding}, and the same concept has been applied to the study of local excitations in 
correlated NiO \cite{larson2007,lee2010} and LiF \cite{lee2013,abbamonte2008}.

The above effective $d^4$ picture, while ideal to study the Jahn-Teller instability and other local properties like the local magnetic moment 
and local excitations, is not suitable for studying long-range properties.
This is because the wave nature of the GaN orbitals, after being integrated out, generates effective magnetic couplings that are 
impurity-configuration dependent between the $\widetilde{\tx{Mn}}$ WOs at different sites. For instance, the magnetic coupling does not only 
depend on the distance between pairs of Mn impurities \cite{Sato10}, but also on the position of other nearby Mn impurities, corresponding to 
three-body and four-body interactions~\cite{x_cui_07}.

Therefore, we proceed to derive an effective $d^5$ picture suitable for studying long-range properties, by integrating out charge fluctuation 
involving Mn-$d$ and Ga orbitals in the multiorbital Anderson Hamiltonian, leaving only the doped hole in the antibonding WOs with 
primarily N-$sp^3$ character. From this we obtain a spin-fermion Hamiltonian with a few 
novel physical effects: $H_\tx{eff} = H_0 + \Delta$, where
\begin{align}\label{h0}
H_0 	&= \sum_{\substack{\bi\bi' mm'\sg}} t^{mm'}_{\bi\bi'} 
	       c^{\dagger}_{\bi m\sg}  c_{\bi' m'\sg} + \ H.c. \
\end{align}
is the Hamiltonian of pure GaN, and
\begin{align}\label{delta}
\Delta 	&= \sum_{\substack{\bj\bi\bi' mm'\sg}} T^{mm'}_{\bj\bi\bi'} 
	       c^{\dagger}_{\bi m\sg}  c_{\bi' m'\sg} \notag\\
	    &+ \sum_{\substack{\bj\bi\bi' mm'\\ \sg\sg'}} J^{mm'}_{\bj\bi\bi'}
	       c^{\dagger}_{\bi m\sg} \boldsymbol \tau_{\sg\sg'}  c_{\bi' m'\sg'} \cdot 
	      \boldsymbol {\hat S}_\bj\ +\ H.c. \ 
\end{align}
contains the influence of the (Ga,Mn) substitution at the primitive unit cell $\bj$, and is thus referred to as the impurity potential.
As usual, $ c_{\bi m\sg}$ ($ c^{\dagger}_{\bi m\sg}$) annihilates (creates) an electron with spin $\sg$ at unit cell $\bi$ in the $m$th WOs.
$t^{mm'}_{\bi\bi'}$ contains the orbital energy (when $\bi = \bi'$ and $m = m'$) and hopping integral of the effective 
$\widetilde{\tx{N}}$-$sp^3$ WOs.
$T^{mm'}_{\bj\bi\bi'}$ and $J^{mm'}_{\bj\bi\bi'}$ represent spin-independent and spin-dependent impurity potentials, respectively.
$\boldsymbol {\hat S}_\bj$ and $\boldsymbol \tau_{\sg\sg'}$ are the spin-$\frac{5}{2}$ unit vector and elements of the Pauli's matrices, 
respectively, and $H.c.$ denotes the Hermitian conjugate. 
To get a better understanding of the origin of this generalized spin-fermion model we illustrate the derivation of the impurity potentials 
from perturbation theory using a simple model in the Supplemental Material \cite{suppl_gamnn_cstate}.

Note that the four WOs with the same unit-cell index are defined to be the $\widetilde{\tx{N}}$-$sp^3$ WOs pointing toward the central 
Ga/Mn ion, one from each surrounding N ion.
With the help of symmetry considerations~\cite{suppl_gamnn_cstate} we choose the proper WOs' subspace corresponding to integrating out the Mn 
and Ga orbitals. These WOs can be constructed from our DFT results within the energy range $[-18.0,~0.4]$ eV, as shown in Figs. \ref{fig:wan_orb} (b) and (c).
In their hybridization tails, one observes clearly bonding with Ga-$sp^3$ [Fig. \ref{fig:wan_orb}(b)] and antibonding with Mn-$d$ 
[Fig. \ref{fig:wan_orb}(c)]. 

Having these WOs at hand, we can then represent the relevant part of the DFT self-consistent Hamiltonian and collect its term into the form of 
Eqs.~(\ref{h0}) and (\ref{delta}). Since this is a faithful representation of the relevant components of the DFT Hamiltonian, its validity is 
actually beyond the second order in the atomic hopping integral. 
A few leading parameters 
in our results are given in Table \ref{tab1}. As expected, they show a rapid decay with the distance from the impurity site.

{\renewcommand{\arraystretch}{1.5}
\begin{table}[h!]
\caption{\label{tab1}
Leading parameters in the impurity potential in meV near the impurity site $\bj$.
NN($\bj$) and NNN($\bj$) denote nearest neighboring and next nearest neighboring sites.
Here, $m\neq m'$.
}
\begin{tabular}{c|c|c|c|c|}
\cline{2-5}
 & $T^{mm}_{\bj\bj\bi'}$ & $T^{mm'}_{\bj\bj\bi'}$ & $J^{mm}_{\bj\bj\bi'}$ & $J^{mm'}_{\bj\bj\bi'}$ \\
 \hline
 \multicolumn{1}{|c|}{$\bi'=\bj$}  & $2488$ & $-170$ & $1752$ & $-633$ \\ \hline
 \multicolumn{1}{|c|}{$\bi'=$ NN($\bj$)}  & $406$  & $885$ & $449$ & $800$ \\ \hline
 \multicolumn{1}{|c|}{$\bi'=$ NNN($\bj$)} & $15$   & $68$ & $< 10$ & $38$ \\ \hline
\end{tabular}
\label{pars}
\end{table}
}

\begin{figure}[htb]
\includegraphics[width=0.5\textwidth,clip=true]{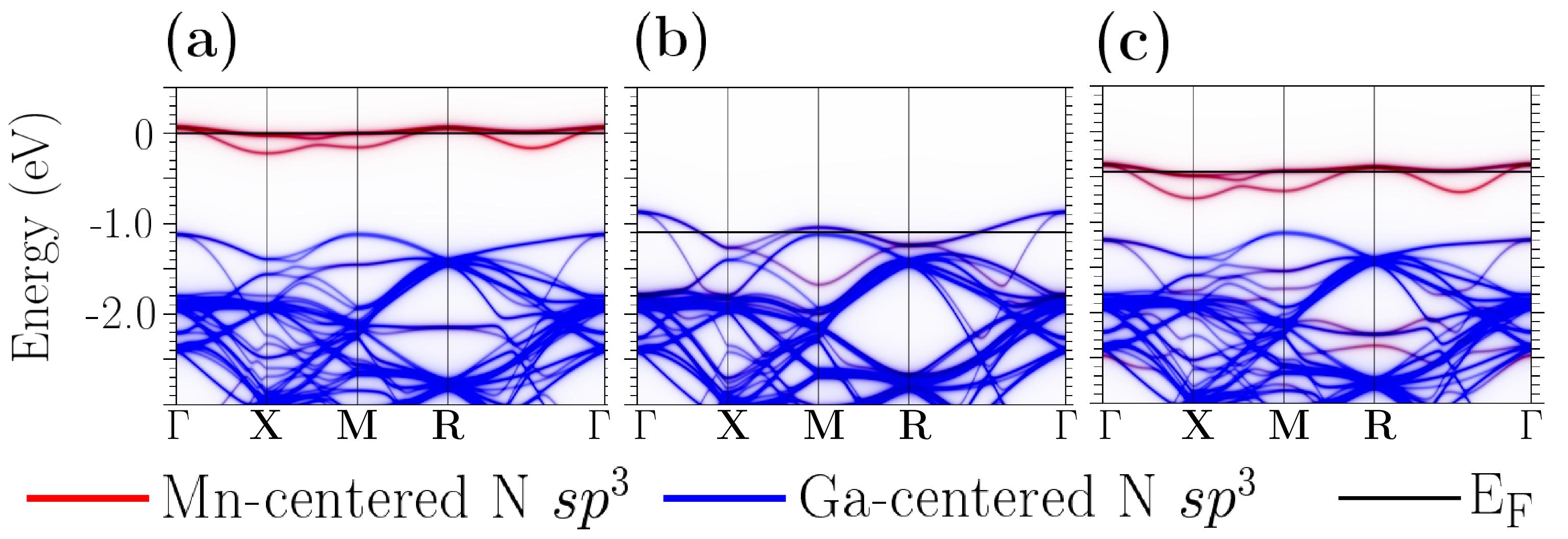}
\caption{
(color online) The spin-majority band structure of Ga$_{31}$MnN$_{32}$ (a) with complete parameters, (b) with the leading orbital energy shift 
$T^{mm}_{\bj\bj\bj}=2488$ meV removed,
and (c) with the two leading exchange-assisted parameters $J^{m m'\neq m}_{\bj\bj\bj}=-633$ and 
$J^{m m'\neq m}_{\bj\bj\bi'}=800$ meV with $\bi'=$ NN($\bj$) removed.}
\label{fig:bands}
\end{figure}

Interestingly, our results reveal a few new physical effects on the carriers besides the previously proposed~\cite{m_abolfath_01} 
antiferromagnetic exchange with the local moment ($J^{mm}_{\bj\bj\bj}=1752$ meV in Tabel~\ref{tab1}). First, the impurity potential contains 
a strong shift of the orbital energy ($T^{mm}_{\bj\bj\bj}=2488$ meV), even \textit{stronger} than the exchange above. This reflects the distinct 
atomic orbitals of Mn (the impurity) and Ga (the host) being integrated out.
In fact, our test shows that if one were to ignore just this parameter, the impurity level [red bands in Fig.~\ref{fig:bands}(a)] would have dropped 
outside the band gap [c.f. Fig.~\ref{fig:bands}(b)], totally destroying the physical characteristics of the system.
Physically, this large orbital energy shift of course induces a strong impurity scattering and a strong tendency toward Anderson localization 
\cite{p_anderson_58}, affecting the carrier mobility, the activation energy, and almost every other essential physical aspect of a semiconductor, 
in addition to altering the effective magnetic coupling between Mn ions.
Second, our results also show a strong exchange-assisted hopping ($J^{m\neq m'}_{\bj\bj\bi'}=-633$ meV and $800$ meV in Tabel~\ref{tab1} close to the 
impurity site.) Again, Fig.~\ref{fig:bands}(c) shows that ignoring these two terms leads to a much smaller spin-dependent splitting of the impurity 
level. Therefore, they not only add to the above impurity effects but also directly modify the magnetic exchange and ordering of Mn impurities.
Both of these two effects are very strong and comparable in strength to the exchange effect included in previous studies, and thus will need to be 
further investigated in the future.

%

It is useful to remark that our approach of employing multiple pictures in understanding different low-energy properties of a many-body system have been 
used in other strongly correlated materials, for examples, in the manganites and the cuprates.
Specifically for the cuprates, the Zhang-Rice singlet description \cite{f_zhang_88} and Emery-Reiter three-spin polaron description \cite{v_emery_88}
are exactly effective  $d^8$ and $d^9$ pictures, parallel to our $d^4$ and $d^5$.
The $d^8$ approach integrates the oxygen degrees of freedom out, resulting in a reduced local magnetic moment $S=0$, similar to our effective $d^4$ 
picture that absorbs implicitly the GaN orbitals and has a smaller moment $S=2$. On the other hand, the $d^9$ picture integrates the charge fluctuation involving the Cu orbitals out and results on doped holes propagating in O orbitals that are correlated antiferromagnetically with the surrounding Cu $S=1/2$ spins,
similar to our effective $d^5$ picture in which carriers live in effective $\widetilde{\tx N}$-$sp^3$ WOs that correlated antiferromagnetically with the Mn $S=5/2$ spins.
Naturally, the more complete $d^9$ picture of cuprates and our $d^5$ picture of (Ga,Mn)N cover a larger energy range than the $d^8$ and $d^4$ pictures respectively, and thus allow richer physical behaviors in general.

To summarize, by investigating the current debate on the Mn valence in Ga$_{1-x}$Mn$_x$N, we advocate three general points in correlated materials: 
1) atomic or ionic valence is only meaningful for high-energy properties but is not very relevant to the low-energy physical properties; 
2) it is often possible to derive multiple effective pictures by integrating out the less relevant degrees of freedom; and
3) for challenging correlated systems, one thus should take advantage of such flexibility and employ the most convenient picture for describing the physical properties of interest.
Specifically, we found the Mn valence of $2+$, but with a nonatomic spin of 4$\mu_B$; illustrating the inadequacy of ionic valence in an atomic picture.
We then demonstrate the feasibility of an effective $d^4$ picture (naturally with $S=4\mu_B$) suitable for studying local instabilities and excitations.
In addition, we derive an effective $d^5$ approach that can be used for future studies of long-range magnetic order, nonlocal magnetic correlation, and other transport properties.
Particularly, our $d^5$ model demonstrates a few novel physical effects beyond previous considerations in the field.
Our results clarify the intrinsic dual nature of the doped holes in the DMS and pave the way for future realistic studies of the magnetism in these systems.
Our study not only resolves one of the outstanding key puzzles in the field, but also emphasizes the general need for multiple effective pictures to describe the rich low-energy physics in many-body systems in general.

We thank P. Derosa for useful feedback on our Letter. This work is supported by NSF DMR-1237565 and NSF EPSCoR Cooperative Agreement No. EPS-1003897 with 
additional support from the Louisiana Board of Regents. W. K. and T. B. were supported by DOE CMCSN DE-AC02-98CH10886.  
T. B. also acknowledges additional support from the Wigner Fellowship of Oak Ridge National Laboratory.
Work by T. B. was partly performed at the Center for Nanophase Materials Sciences, a DOE Office of Science user facility.
Supercomputer support is provided by the Louisiana Optical Network Initiative (LONI) and HPC@LSU computing resources.
\bibliographystyle{apsrev4-1}
\bibliography{master}

\begin{widetext}
\vspace{0.4in}

\setcounter{figure}{0}
\setcounter{equation}{0}
\renewcommand{\figurename}{Figure S\!\!}
\interfootnotelinepenalty=10000

\section*{\Large\bf Supplementary Information}\label{suppl_gamnn_cstate}

The following derivation is to illustrate how ``high-energy'' degrees of freedom in the Ga$_{1-x}$Mn$_x$N system are integrated out while the ``low-energy'' ones are retained to form the effective low-energy potentials, i.e. the spin-independent \& -dependent potentials. To this end we perform $2^\tx{nd}$ order perturbation theory on a simple model Hamiltonian consisting of 5 Mn-$d$ orbitals and 4 N-$sp^3$ orbitals pointing towards the Mn.
The results of this calculation should, of course, not be taken too literally. 
Instead, they merely serve the purpose of providing additional understanding of the first principles results of the manuscript.

The Hamiltonian is given by:
\begin{align}\label{hbare}
 H_\tx{bare} &= \sum_{mm'\sg} \l(t^{mm'}\cp_{ m\sg} c_{m'\sg} + h.c.\r) 
      + \sum_{l \sg} \vep_{l } n_{l \sg}^d\ns
      &+ U \sum_{l } n_{l \upa}^d n_{l \dna}^d
      + U' \sum_{l < l' \sg\sg'} n_{l \sg}^d n_{l' \sg'}^d 
      - J \sum_{l < l' \sg\sg'} \dd_{l \sg}\dd_{l' \sg'}d_{l' \sg}d_{l \sg'}\ns
      &+ \sum_{\langle m, l \rangle  \sg} \l(V_{\langle m,l \rangle}\cp_{m\sg} d_{l \sg} + h.c.\r)
\end{align}
where,
\begin{align}
 m, m' &: \tx{indices of N-$sp^3$ orbitals pointing toward Mn},\ns
 l , l'  &: \tx{indices of Mn-$d$ orbitals},\ns
 \sg,\sg' &: \tx{spin indices},\ns
 t^{mm'} &: \tx{($\equiv t$) N-$sp^3$ to N-$sp^3$ hopping parameter for $m \neq m'$},\ns
 t^{mm'} &: \tx{($\equiv\vep_m$) on-site energy of N-$sp^3$ orbitals for $m = m'$},\ns 
 \vep_{l } &: \tx{on-site energy of Mn-$d$ orbitals},\ns
 U &: \tx{intra-orbital Coulomb interaction of the Mn-$d$ orbitals},\ns
 U' &: \tx{inter-orbital Coulomb interaction of the Mn-$d$ orbitals},\ns
 J &: \tx{Hund's coupling of Mn-$d$ orbitals},\ns
 V_{\langle m,l \rangle} &: \tx{hybridization parameter},\ns
\cp, c &: \tx{creation and annihilation operators of N-$sp^3$ electrons},\ns
\dd, d &: \tx{creation and annihilation operators of Mn electrons},\ns
n &: \cp c,\ns
n^d &: \dd d.\notag
\end{align}
Moreover, due to the size of the Hund's coupling $J$ we assume that the spin of the $e_g$ electrons 
is always aligned with the one of the $t_{2g}$ electrons. 
Finally, we specifically deal with a system with only one doped hole.

\begin{figure}[ht]
\centering
\includegraphics[width=1\textwidth,clip=true]{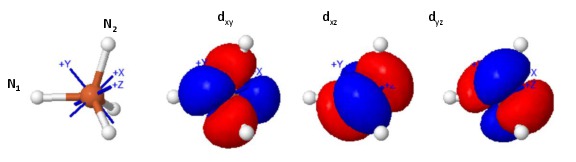}
\caption{
Model of Mn-$t_{2g}$ orbitals in tetrahedral crystal field of surrounding N atoms obtained from \protect\url{http://wwwchem.uwimona.edu.jm:1104/courses/CFT_Orbs.html}.
}\label{tetra}
\end{figure}

Now consider the Hamiltonian (\ref{hbare}) when the hybridization is turned-off 
($V_{\langle m,t_{2g} \rangle}$ $=0$). 
In this particular situation we can group all states into two disconnected Hilbert spaces: low- and high-energy subspaces.
The low-energy Hilbert space contains a hole in the N-$sp^3$ orbitals while the Mn-$t_{2g}$ and $e_{g}$  orbitals are 
half-filled and in the high-spin configuration ($S=\frac{5}{2}$) with energy 
$\displaystyle E_{d^5} = 3\vep_m + 5\vep_l + \frac{5!}{2!3!} \l(U'-J\r)$, 
where factor $3$ comes from the number of occupied N orbitals around Mn, 
the factor $5$ counts the number of occupied Mn-$d$ orbitals, 
and the factor of $10=5!/(2!\,3!)$ is the number of Mn-$d$ orbitals pairs.
Meanwhile, the rest of the states are part of the high-energy Hilbert space;
in particular the atomic $d^4$ states with a hole localized on the 
Mn site and energy $E_{d^4} =  4\vep_m+ 4\vep_l + 6 \l(U'-J\r)$, where the factor of $6=4!/(2!\,2!)$ is the number of Mn-$d$  pairs, 
belongs to the high-energy sector. 
The energy difference between the high- ($d^4$) and low-energy ($d^5$) Hilbert spaces becomes
$\Delta_E = E_{d^4} - E_{d^5} = \vep_m - \vep_l - 4\l( U' - J \r)$.
When the hybridization is turned on ($V_{\langle m,t_{2g} \rangle}\neq0$) 
the low- and high-energy states are mixed. The task is then to extract the change in the low-energy Hilbert 
space due to the hybridization between $d^4$ \& $d^5$ degrees of freedom. 

The change in the energy of $d^5$ to $2^\tx{nd}$ order in perturbation theory is given by 
\footnote{K. Gottfried and T.-M. Yan, {\it Quantum Mechanics: Fundamentals, $2^\tx{nd}$ ed.}, 2003.}
\begin{equation}\label{heff}
\langle\beta|\Delta|\alpha\rangle = \langle\beta|\hat V|\alpha\rangle - \sum_{\mu}
\frac {\langle\beta|\hat V|\mu\rangle\langle\mu|\hat V|\alpha\rangle} {\Delta_E}  \ \ce
\end{equation}
where 
\begin{align}
|\alpha\rangle, |\beta\rangle &: \tx{states in the low-energy $d^5$ Hilbert space},\ns
|\mu\rangle &: \tx{states in the high-energy $d^4$ Hilbert space},\ns
\hat V &: \tx{the hybridization operator $=\sum_{ml \sg} 
	\l(V_{\langle m,t_{2g} \rangle}\cp_{m\sg} d_{l \sg} + h.c.\r)$},\ns
\Delta_E &: \tx{energy difference between the high- ($d^4$) and low-energy ($d^5$) Hilbert space.}
\end{align}
Furthermore, because the hybridization only connects the low-and high-energy terms the first term in Eq.~(\ref{heff})
vanishes.

We particularly want to integrate out the atomic $d^4$ states ($|\mu\rangle$ states) and
keep the effective $d^5$ states ($|\alpha\rangle$, $|\beta\rangle$).
The $d^4$ states are formed when the hole hops to the Mn-$t_{2g}$ orbitals whereas the hole is located effectively
in the N-$sp^3$ orbitals for the $d^5$ states. If $| m \sg,m_s\rangle$ denote $|\alpha\rangle$ 
and $|\beta\rangle$; i.e. representing orbital ($m$), and spin ($\sg$) state of the hole 
and the projected spin state of Mn ($m_s$), we can group the different components of 
the second term of Eq.~(\ref{heff}) into four contributions, all of them involving virtual hopping to
Mn-$t_{2g}$ orbitals (virtually forming a $d^4$ state):
\begin{enumerate}[(a)]
 \item low-energy states in which the hole stays in the same N-$sp^3$ orbital without exchanging its spin with Mn;
      i.e. $\displaystyle \langle\beta|\Delta|\alpha\rangle = \sum_{\mu} \frac {\langle  m\sg,m_s|\hat V|\mu\rangle
	    \langle\mu|\hat V| m\sg,m_s\rangle} {\Delta_E} $,\label{t1}
 \item low-energy states in which the hole stays in the same N-$sp^3$ orbital but exchanges its spin with Mn;
      i.e. $\displaystyle  \langle\beta|\Delta|\alpha\rangle = \sum_{\mu} \frac {\langle  m\sg',m'_s|\hat V|\mu\rangle
	    \langle\mu|\hat V| m\sg,m_s\rangle} {\Delta_E} $,\label{t2}
 \item low-energy states in which the hole hops from one N-$sp^3$ to another N-$sp^3$ without spin exchange;
      i.e. \\$\displaystyle \langle\beta|\Delta|\alpha\rangle = \sum_{\mu} \frac {\langle  m'\sg,m_s|\hat V|\mu\rangle
	    \langle\mu|\hat V| m\sg,m_s\rangle} {\Delta_E} $,\label{t3}
 \item low-energy states in which the hole hops from one N-$sp^3$ to another N-$sp^3$ with spin exchange;
      i.e. \\$\displaystyle \langle\beta|\Delta|\alpha\rangle = \sum_{\mu} \frac {\langle  m'\sg',m'_s|\hat V|\mu\rangle
	    \langle\mu|\hat V| m\sg,m_s\rangle} {\Delta_E} $.\label{t4}
\end{enumerate}
Furthermore, we notice there are three possible Mn-$t_{2g}$ orbitals where the hole
can hop from an initial N-$sp^3$ orbital to form a virtual $d^4$ state ($|\mu\rangle$).
For (\ref{t1}) and (\ref{t2}) these three possibilities give a factor of $3$ to the calculation.
However, they only give a factor of 1 for (\ref{t3}) and (\ref{t4}) due to the opposite sign that 
some of the hybridization factors have. 
For instance, Fig. S\ref{tetra} shows that the hopping from N$_1$-$sp^3$ to N$_2$-$sp^3$ via $d_{xz}$ will have an opposite sign of those via $d_{xy}$ and $d_{yz}$, since the lobes pointing towards N$_1$ and N$_2$ are both positive for $d_{xz}$, while they change sign for $d_{xy}$ and $d_{yz}$.
\begin{align}
    \langle \tx{N}_1 sp^3_a \sg',m'_s|\hat V|d^4_{xy}\rangle 
    \langle d^4_{xy}|\hat V|\tx{N}_2 sp^3_b \sg,m_s\rangle 
 &=
   -\langle \tx{N}_1 sp^3_a \sg',m'_s|\hat V|d^4_{xz}\rangle 
    \langle d^4_{xz}|\hat V|\tx{N}_2 sp^3_b \sg,m_s\rangle\ns 
 &=\quad\!
    \langle \tx{N}_1 sp^3_a\sg',m'_s|\hat V|d^4_{yz}\rangle 
    \langle d^4_{yz}|\hat V|\tx{N}_2 sp^3_b \sg,m_s\rangle, 
\end{align}
where $|d^4_{\nu}\rangle$ denotes a particular $t_{2g}$ orbital where the hole virtually hops into.

Next we show that by calculating some elements of
\begin{equation}\label{deltao} 
\Delta = - \sum_{\mu} \frac {\hat V|\mu\rangle\langle\mu|\hat V} {\Delta_E}, 
\end{equation}
we can compactly express $\Delta$ using the second-quantized form:
\begin{align}\label{deltasq}
\Delta 	= \sum_{\substack{ mm'\sg}} T^{mm'}
	       \tilde c^{\dagger}_{ m\sg}  \tilde c_{ m'\sg} 
	    + \sum_{\substack{ mm'\\ \sg\sg'}} J^{mm'}
	       \tilde c^{\dagger}_{ m\sg} \boldsymbol \tau_{\sg\sg'}  \tilde c_{ m'\sg'} \cdot 
	      \boldsymbol S\ +\ h.c. \ce
\end{align}
Here $\boldsymbol S$, $\boldsymbol \tau_{\sg\sg'}$,  and $\tilde c^{\dagger}_{ m\sg}$ ($\tilde c_{ m\sg}$)
are, respectively, the quantum Mn spin-$\frac{5}{2}$ vector located at the origin, the Pauli matrices, and
the creation (annihilation) operator of quasiparticles. 
Given the large moment of the Mn spin it can be approximated as being a classical vector for practical applications.
In order to fix the parameters of the effective Hamiltonian in Eq. (\ref{deltasq}) in terms of the parameters of the bare Hamiltonian, 
we need to evaluate Eq. (\ref{deltao}) for four different pairs of $|\alpha\rangle$ and $|\beta\rangle$.
\begin{enumerate}[A.]
  \item To get $T^{mm}$ and $J^{mm}$:\label{acal}
  \begin{enumerate}[I.]
    \item $|\alpha\rangle = |\beta\rangle = | m\sg= \frac{1}{2},m_s = \frac{5}{2}\rangle$:\label{ae1}\\ \\
      Since Pauli's principle does not allow two identical particles with the same spin to be at the same site, 
      applying $\hat V$ on $|\alpha\rangle$ or $|\beta\rangle$ will result in zero, 
      thus from Eq. (\ref{deltao}) we have $\langle\beta|\Delta|\alpha\rangle = 0$. 
      Eq. (\ref{deltasq}), on the other hand, gives $\langle\beta|\Delta|\alpha\rangle = T^{mm} + \frac{5}{2} J^{mm}$.\\ 
    \item $|\alpha\rangle = |\beta\rangle = | m\sg= -\frac{1}{2},m_s = \frac{5}{2}\rangle$:\label{ae2}\\ \\
    In this case, since the hole in a given N-$sp^3$ orbital has opposite spin to the ones in Mn-$d$, 
    applying $\hat V$ on $|\alpha\rangle$ or $|\beta\rangle$ will allow the hole to hop from the N-$sp^3$ orbital
    to a particular Mn-$d$ orbital and back to the same N-$sp^3$ orbital. 
    These hopping processes, furthermore, interfere constructively giving rise to a factor of 3.
    Hence, Eq. (\ref{deltao}) yields $\langle\beta|\Delta|\alpha\rangle = -3\frac{\left|V_{\langle m,t_{2g} \rangle}\right|^2}{\Delta_E}$. 
	Eq. (\ref{deltasq}), furthermore, gives  $\langle\beta|\Delta|\alpha\rangle = T^{mm} - \frac{5}{2} J^{mm}$,
    \suspend{enumerate}
    thus, from \ref{ae1} and \ref{ae2} we get $T^{mm} = -\frac{3|V_{\langle m,t_{2g} \rangle}|^2}{2\Delta_E}$ and 
    $J^{mm} = \frac{3|V_{\langle m,t_{2g} \rangle}|^2}{5\Delta_E}$. 

  \item To get $T^{m\neq m'}$ and $J^{m\neq m'}$ (here, for instance, $m = sp^3_{N_1}$ and $m' = sp^3_{N_2}$):
  \begin{enumerate}[I.]
    \item $|\alpha\rangle = | m\sg= \frac{1}{2},m_s = \frac{5}{2}\rangle,\ 
	 |\beta\rangle = | m'\sg= \frac{1}{2},m_s = \frac{5}{2}\rangle$:\label{be1}\\ \\
	 For the same reason as in  \ref{ae1}, applying $\hat V$ on $|\alpha\rangle$ or $|\beta\rangle$ will result in zero, 
	 thus from Eq. (\ref{deltao}) we have $\langle\beta|\Delta|\alpha\rangle = 0$. 
	 Whereas, Eq. (\ref{deltasq}) gives $\langle\beta|\Delta|\alpha\rangle = T^{m\ne m'} + \frac{5}{2} J^{m\ne m'}$.\\
      
    \item $|\alpha\rangle = | m\sg= -\frac{1}{2},m_s = \frac{5}{2}\rangle,\ 
	 |\beta\rangle = | m'\sg= -\frac{1}{2},m_s = \frac{5}{2}\rangle$:\label{be2}\\ \\
	 Like \ref{ae2}, applying $\hat V$ on $|\alpha\rangle$ or $|\beta\rangle$ will make the hole to hop from a given N-$sp^3$ orbital 
    to a particular Mn-$d$ orbital but hop back to a different N-$sp^3$ orbital. 
    However, unlike \ref{ae2}, some hopping processes, interfere destructively giving rise to a factor of 1 instead of 3.
    Hence, Eq. (\ref{deltao}) yields 
    $\langle\beta|\Delta|\alpha\rangle = -\frac{\l(V_{\langle m,t_{2g} \rangle} \cdot V_{\langle t_{2g},m'\rangle}\r)}{\Delta_E}$. 
    Whereas Eq. (\ref{deltasq}) gives $\langle\beta|\Delta|\alpha\rangle = T^{m\ne m'} - \frac{5}{2} J^{m\ne m'}$\\
    \suspend{enumerate}
    Accordingly, from \ref{be1} and \ref{be2} we get $T^{m\neq m'} = 
    -\frac{\l(V_{\langle m,t_{2g} \rangle} \cdot V_{\langle t_{2g},m'\rangle}\r)} {2\Delta_E}$ and $J^{m\ne m'} = 
    \frac{\l(V_{\langle m,t_{2g} \rangle} \cdot V_{\langle t_{2g},m'\rangle}\r)}{5\Delta_E}$. 
\end{enumerate}

To summarize we have illustrated the effective low-energy $d^5$ model, specifically, we show within 
the $2^\tx{nd}$ order perturbation theory scheme how the spin-dependent and spin-independent potentials intuitively emerge by virtually
exchanging a hole between the $d^5$ and $d^4$ states.

\end{widetext}

\end{document}